# Design Automation for Gray-Code Quantum Read-Only Memory

Hao-Yu Lu
*Intelligent Computing Industrial Doctorate Program*
*Miin Wu School of Computing*
*National Cheng Kung University*
Tainan, Taiwan
np8121595@gs.ncku.edu.tw

Yu-Ting Kao
*Electronic and Optoelectronic System Research Laboratories*
*Industrial Technology Research Institute*
Hsinchu, Taiwan
SunnyKao@itri.org.tw

Yeong-Jar Chang
*Electronic and Optoelectronic System Research Laboratories*
*Industrial Technology Research Institute*
Hsinchu, Taiwan
ot@itri.org.tw

Chao-Hung Wang
*Miin Wu School of Computing*
*National Cheng Kung University*
Tainan, Taiwan
chwang@gs.ncku.edu.tw

Darsen D. Lu
*Department of Electrical Engineering*
*National Cheng Kung University*
Tainan, Taiwan
darsenlu@mail.ncku.edu.tw

***Abstract*—We propose a programmable quantum read-only memory based on Gray code encoding, termed GQROM, designed to efficiently load classical data into quantum circuits. This architecture mitigates critical bottlenecks in near-term quantum computing by reducing initialization overhead and gate-induced errors. By utilizing Gray-code, where consecutive addresses differ by only a single bit, GQROM minimizes the transition complexity between states. We demonstrate that architecture's flexibility in accessing targeted data via initial state modification. The entire design, including a systematic EDA flow, was implemented and validated on the Qiskit platform, confirming its practical feasibility. These results underscore GQROM's potential as a scalable and hardware-friendly solution for efficient state preparation in quantum circuits.**

***Index Terms*—Quantum Computing, Read-only Memory, EDA, Gray code, NISQ**

## I. Introduction

In classical computing, read-only memory (ROM) is a foundational component that provides essential non-volatile storage. This permanent data retention makes it the ideal solution for storing firmware and boot loaders. ROM is also widely used to implement lookup tables (LUTs), which store pre-calculated values for complex mathematical functions or data conversions. Although technology has evolved into modern variations—like programmable ROM (PROM), erasable ROM (EPROM), and electrically erasable ROM (EEPROM). The core purpose remains: to provide a stable, persistent, and efficient data store that is fundamental to the architecture and operation of all classical processors.

In the quantum computing realm, while most ongoing research has prioritized the development of quantum processing units (QPUs), current architectures remain critically bottlenecked by their reliance on repeated quantum-classical transfers for data storage. This approach introduces significant latency and drastically increases susceptibility to decoherence and transmission errors [1] [2]. Achieving genuinely efficient computation, therefore, necessitates realizing memory functions directly within the quantum circuit domain. This is fundamentally challenging due to the non-cloning nature of qubits, which prevents direct duplication of quantum states.

Even so, functions like quantum memory (QM) are still required to achieve practical, large-scale operation [3]. Without them, computations are fundamentally constrained by the fragile coherence times of processing qubits and the overhead of data movement between quantum and classical domains [4]–[6]. Quantum read-only memory (QROM), a special QM with ROM function, addresses the bottleneck of loading classical data into quantum circuits as initial states. QROM encodes classical information into a quantum-accessible structure that can be efficiently queried using quantum address registers [7]. This mechanism is essential for algorithms that require complex input states or the repeated use of stored data, as it replaces costly, error-prone state preparation with a stable and resource-efficient query [8], [9]. Developing reliable, low-overhead QROM is therefore as critical as advancing quantum logic gates for building universal quantum computers.

This work introduces a novel and resource-optimized QROM design flow, termed GQROM, built entirely from standard quantum logic gates. Our primary contribution is the integration of Gray code-based encoding with programmable QROM architecture. This combination is designed to significantly reduce the required quantum gates and overall circuit depth, leading to higher computational efficiency and lower hardware error rates.

The remainder of this paper is as follows. Section II broadly reviews current QM-related technologies and why QM is needed for quantum computers. Section III introduces the basic building blocks of GQROM, focusing on the mathematical justification and resource benefits of the Gray code encoding scheme. Section IV details the algorithm for GQROM synthesis and the methodology for data loading. In Section V, we present a thorough complexity analysis comparing our approach against conventional implementation. Finally, Section VI concludes the paper.

**TABLE I**. COMPARISON OF QUANTUM MEMORY TECHNOLOGIES

| Quantum Memory | Primary Function | Typical Applications | Strengths | Limitations |
|---|---|---|---|---|
| **Quantum Repeater** | Stores and preserves entanglement across long distances | Quantum communication, quantum internet, distributed quantum computing | Extends range of entanglement distribution, enables fault-tolerant networking | Requires long-lived memory qubits, more relevant to communication than local computing |
| **QRAM** | Allows superposition queries to large classical datasets | Quantum machine learning, Grover search, big-data algorithms | Enables scalable data access, supports flexible addressing | High hardware complexity, error-prone in near-term devices |
| **QROM** | Encodes classical data into quantum-accessible structures for efficient querying | Hamiltonian simulation, quantum arithmetic, fixed-data algorithms, look-up table | Low gate depth with optimizations, resource-efficient, avoids repeated quantum-classical transfers | Read-only, not suitable for dynamic or large-scale data loading |

## II. TECHNOLOGIES OF QUANTUM MEMORY

An integrated QM provides a viable alternative. By encoding classical data into quantum-accessible forms, QM allows QPUs to handle large datasets with the massive parallelism inherent to quantum operations [10] [11]. Critically, compared with classical memory, an optimally designed QM offers the potential for exponential parallel access to N data entries using only O(log N) addressing qubits, significantly expanding computational capacity.

The field of QM features several complementary forms, each targeting distinct challenges (**Table I**). Quantum random-access memory (QRAM) [12] [13], for instance, focuses on flexible superposition queries over large classical datasets, serving as a crucial component for quantum machine learning and Grover-type searches. Quantum repeaters [14], conversely, function as distributed memory nodes to preserve entanglement across long distances, enabling scalable quantum networks. In contrast to these, QROM serves a more specialized but equally vital role: providing fast, structured, and resource-efficient read-only access to fixed datasets within a quantum processor [8] [9]. Thus, while QRAM and quantum repeaters address large-scale data access and communication, QROM offers a lightweight, practical solution for embedding essential constants (e.g., lookup tables, Hamiltonian coefficients) directly into computation circuits [15].

## III. GQROM ARCHITECTURE AND HIGH EFFICIENCY DESIGN

### *A. Core GQROM Circuit Architecture*

**Fig. 1** illustrates a GQROM composed of upper address qubits and lower data qubits. The operation begins with an initial state preparation. The "Initial" part is mainly composed of Hadamard gates or Pauli-x gates. They are used to create a uniform superposition of memory locations or determine the access of particular storage positions. Subsequently, a series of "Store" blocks ($Store_1$ to $Store_4$ shown in **Fig. 1**) encode the data stored in the GQROM. Each block uses multi-controlled X (MCX) gates to write data conditioned on a specific address. The position and number of MCX gates are determined by the "1" s in the stored data in binary form. To target different memory locations, a Gray code addressing scheme is implemented by selectively applying Pauli-X gates to the address qubits. After the iterative mapping process is complete, a final reversal logic gate is applied to the most significant address qubit. This Pauli-X gate is crucial for maintaining the unitarity of quantum states by effectively reversing the address encoding operations and ensuring the validity of the GQROM.

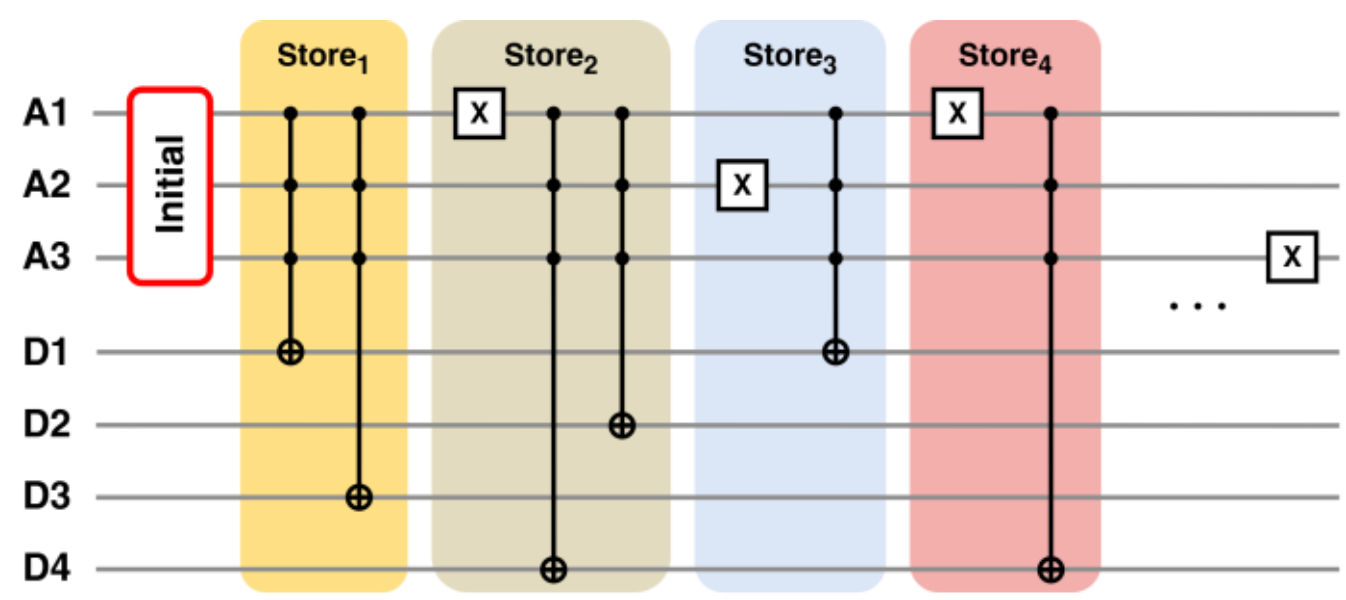

**Fig. 1**. Core GQROM circuit block design

### *B. GQROM Periphery and Purpose*

Circuit peripheries are also important while implementing GQROM, they can not only help process the data output from assigned states but also expand the possible application scenario. GQROM typically functions as a core component within a larger, modular quantum circuit, as illustrated in the system diagram (**Fig. 2**). While its applications are varied, we take a primary use case, which is serving as a quantum look-up table, for example. The functionality of each block in this architecture is detailed below.

- The pre-circuit blocks are responsible for preparing the initial quantum states that will serve as inputs to the GQROM. These circuits can perform complex computations to generate the address and data. Note that pre-circuits are not limited to quantum circuits but can also be classical algorithms.
- The GQROM module is the central look-up table of the system. It takes a quantum state from the pre-circuits as an address and, in a coherent quantum manner, maps it to a corresponding data state.
- Ancilla block: This block provides additional ancilla qubits. These qubits are not part of the primary data but are essential for intermediate calculations within the attached blocks.
- Attached block: This block represents a task-specific quantum operation that utilizes the data retrieved from the GQROM.
- A projective measurement ("M") is performed on specific qubits to collapse their quantum state into a definite classical outcome (0 or 1). This is the standard method for extracting classical information from a quantum computation.
- Alternatively, the quantum state output from the attached block can be directly fed into a post-circuit. This allows for further quantum computation, using the result of the GQROM look-up as an input for subsequent algorithms before any measurement occurs.

This structured, block-based design highlights the modularity of quantum circuit design and clarifies the role of the GQROM as an integrated data-retrieval component within sophisticated quantum algorithms.

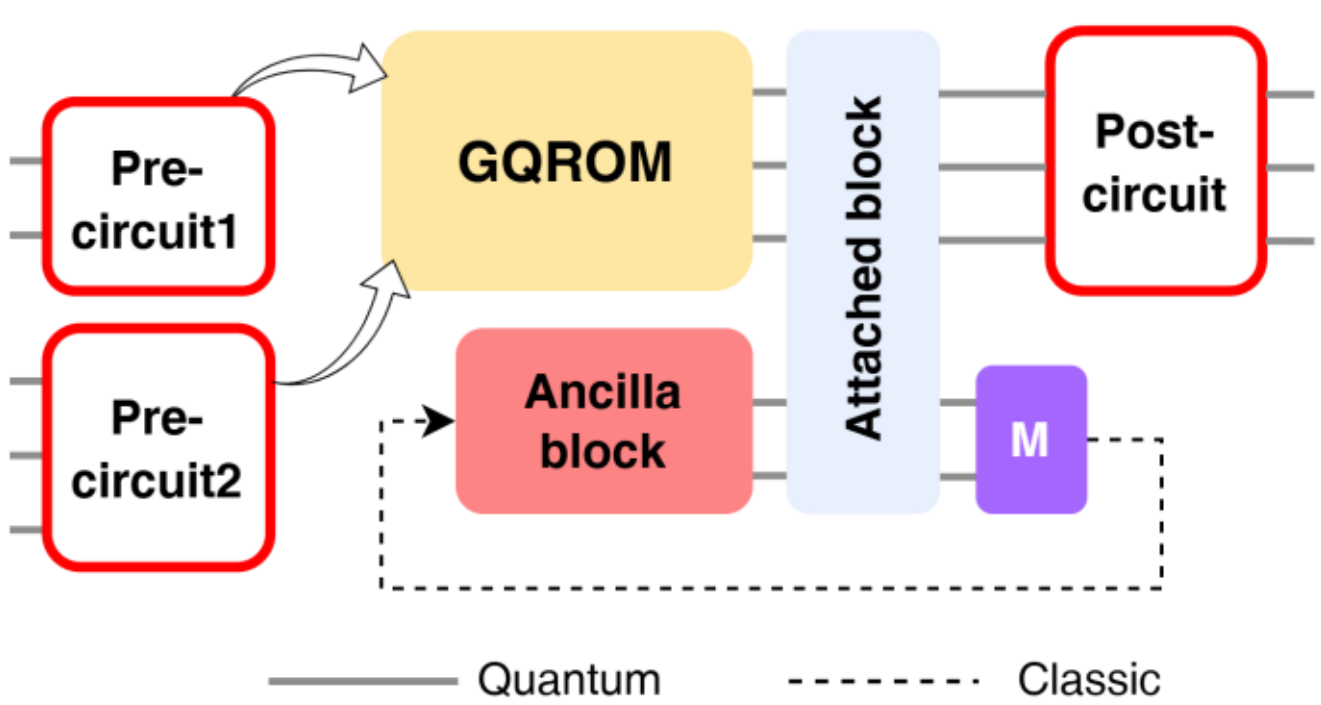

**Fig. 2**. System diagram of GQROM with peripheries

### C. Gray Code Encoding

Gray code is a binary numeral system where two successive values differ in only one bit. This single-step distance property is critical for minimizing errors in systems where state transitions are physical, such as in electromechanical rotary encoders. **Table. II** shows the comparison between decimal code and binary code to Gray code. Gray code eliminates this failure mode by design, ensuring reliable readings. In quantum computing, this principle is leveraged to optimize circuit complexity. By encoding addresses in Gray code, the control logic for accessing consecutive memory locations is simplified, as only a single bit needs to be toggled at a time, thereby reducing the required number of quantum gates.

**TABLE II.** BINARY CODE TO GRAY CODE (TAKE 3-BIT FOR EXAMPLE)

| Store_i | Binary | Data | Gray | Data |
|---|---|---|---|---|
| 1 | 000 | 0 | 000 | 0 |
| 2 | 001 | 1 | 001 | 1 |
| 3 | 010 | 2 | 011 | 3 |
| 4 | 011 | 3 | 010 | 2 |
| 5 | 100 | 4 | 110 | 6 |
| 6 | 101 | 5 | 111 | 7 |
| 7 | 110 | 6 | 101 | 5 |
| 8 | 111 | 7 | 100 | 4 |

## IV. EDA DESIGN FLOW FOR GQROM

To enhance the accessibility of GQROM, we have formalized its design flow into a systematic synthesis procedure, as outlined in the flowchart in **Fig. 3**. This flow provides an automated process that translates a classical dataset into an optimized quantum circuit according to specified address and data registers.

The procedure begins by defining the classical inputs: the data to be stored and the required number of address (n) and data (m) qubits. A critical pre-computation step then generates a Gray code sequence for all $2^n$ addresses and reorders the input data to match the new sequence. The quantum circuit synthesis starts with all qubits initialized to the $|0\rangle$ state. Next, specify the dimension required for computation with different initializations. The core of the algorithm is the data-loading loop, which iterates through the reordered Gray code sequence. For each address, the corresponding data is encoded using MCX gates. Crucially, the transition between consecutive Gray code addresses are efficiently managed by applying a single Pauli-X gate. This single-gate transition is the key to minimizing the circuit's overall gate count and depth. This process culminates in a final, optimized quantum circuit that represents the complete GQROM, efficiently encoding the entire dataset.

## V. FUNCTIONAL VALIDATION OF GQROM

To demonstrate the functionality of the GQROM, we executed a data storage and readout experiment on the IBM Qiskit platform [16], with the results shown in **Fig. 4**. Specific memory locations were accessed by modifying the initial state preparation block (as shown in **Fig. 1**).

For this test, the dataset {3, 7, 8, 2, 4, 6, 1, 5} was encoded into the GQROM for addresses 0 through 7. **Fig. 4(a)** demonstrates a selected query: by applying a Hadamard gate to only the most significant address qubit, a superposition of addresses $|000\rangle$ (decimal 0) and $|100\rangle$ (decimal 4) is created. Consequently, the measured output correctly shows a superposition of their stored data values: 3 (binary 0011) and 4

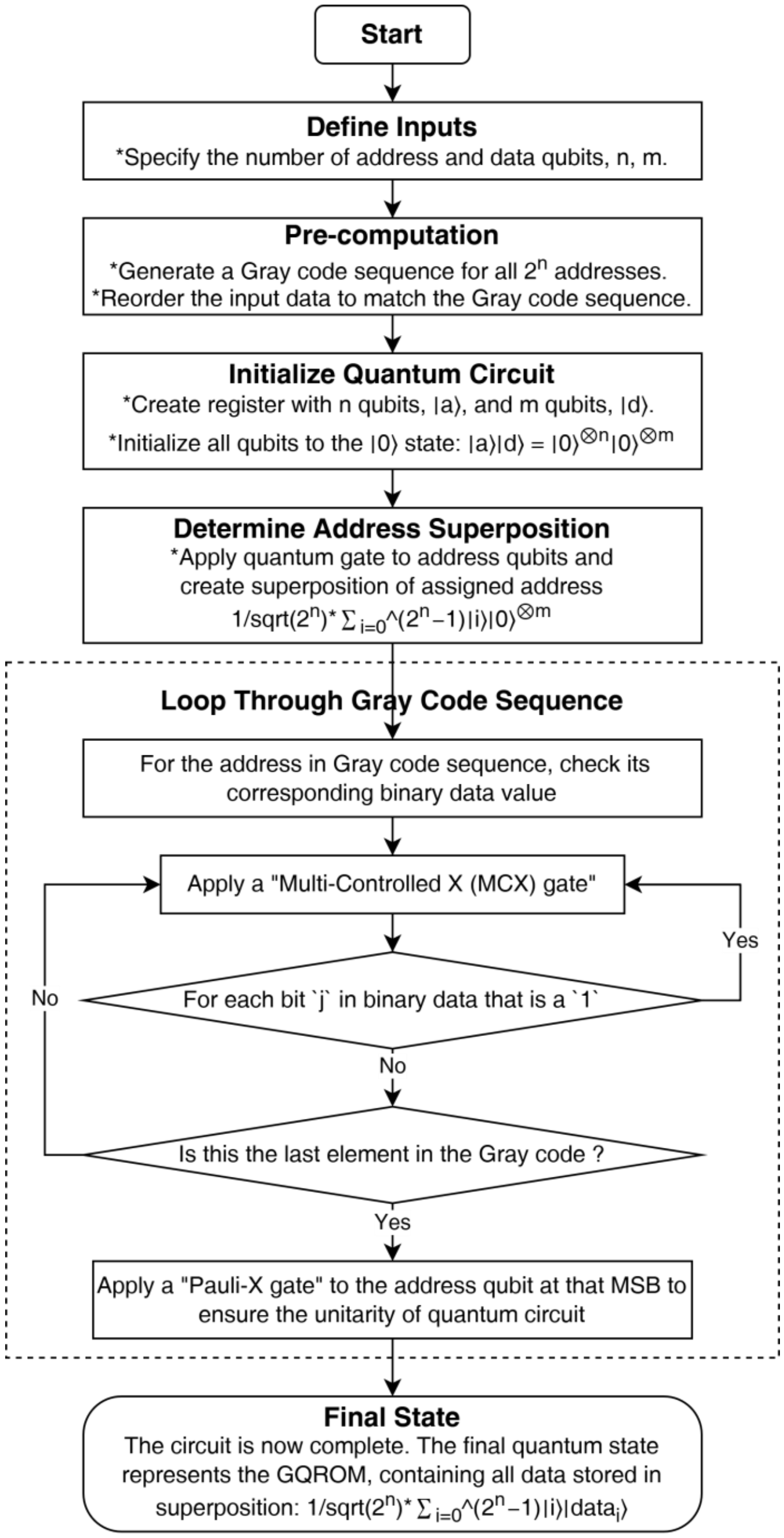

**Fig. 3**. Design Flow chart for GQROM

(binary 0100). Following this principle, the results in **Fig. 4(b)** and (c) validate the GQROM's operation over larger superpositions, showing accurate data retrieval for $2^2$ and $2^3$ states, respectively. Finally, **Fig. 4(d)** demonstrates an alternative initialization using Pauli-X gates, confirming the GQROM's flexibility in accessing specified data to fit different application needs.

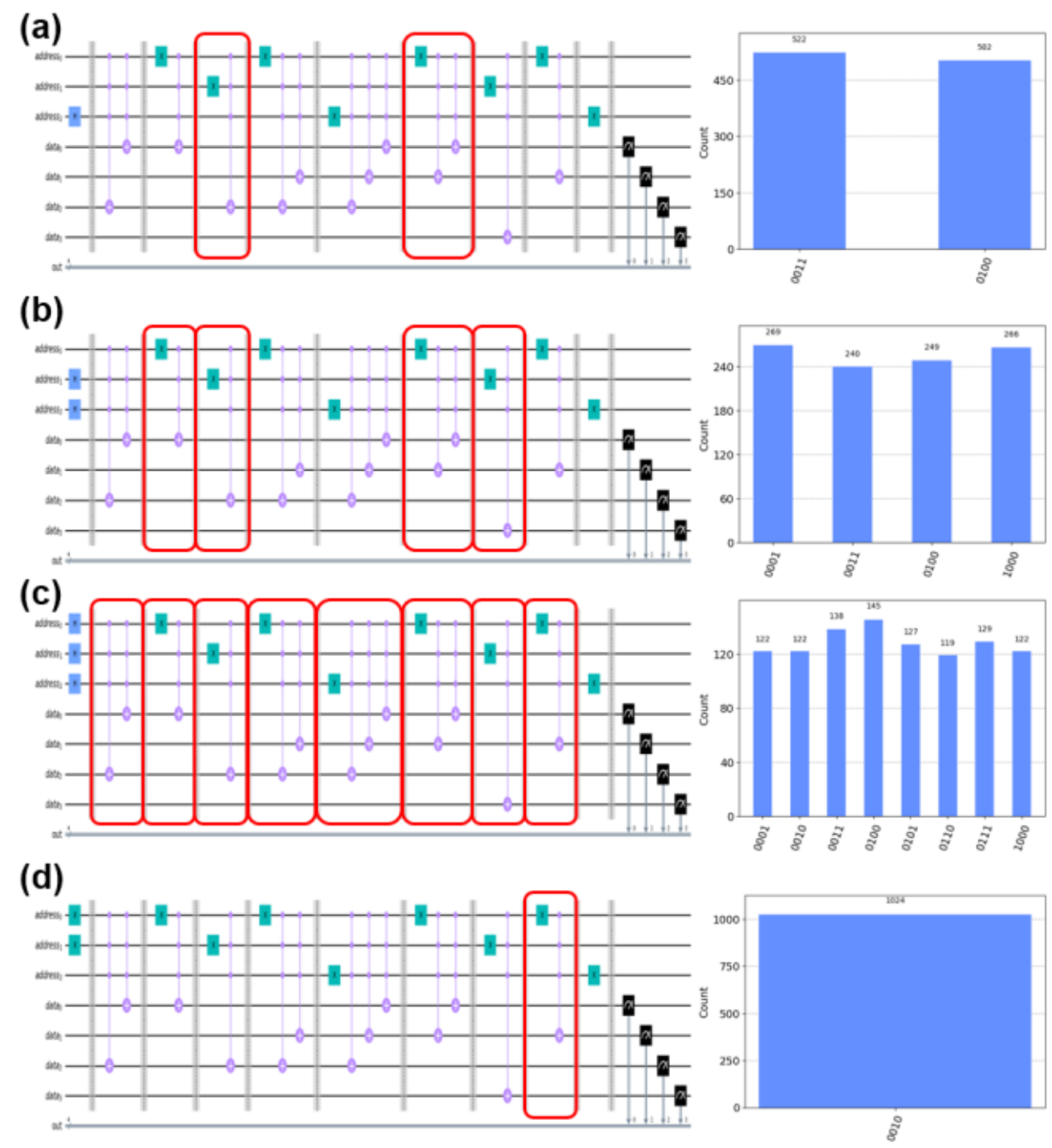

**Fig. 4**. GQROM with different "Initial". (a)(b)(c) 1 or 2 or 3 Hadamard gates are added, respectively. (d) two Pauli-X gate is added. The histograms prove that we can access specified data with different initial setting and create different superposition depending on the applications.

## VI. Conclusion

In this work, we've successfully developed and validated the GQROM, an efficient QROM design with the address encoded using Gray code. We provided a systematic EDA design flow to automate GQROM synthesis, enhancing its accessibility and integration. By demonstrating our design on the widely used Qiskit platform, we have confirmed that GQROM is a feasible and easily integrable tool for complex quantum workflows. The core advantages of our design are efficient gate usage, hardware-friendly encoding, and improved noise resilience. These establish GQROM as a vital component for the current quantum applications, paving the way for more sophisticated and reliable computations on near-term devices.

## Acknowledgment

This work is supported by the Miin Wu School of Computing, NCKU. The authors would like to gratefully acknowledge NCKU Quantum Frontier of Research & Technology (QFort) for quantum computational resources. As well as the financial support from the Department of Industrial Technology, Ministry of Economic Affairs, Taiwan under the "Large Language Model Computing Chip Program" of the Taiwan Chip-based Industrial Innovation Program.